\documentclass[usenatbib]{mn2e}
\usepackage{amsmath, amssymb}
\usepackage{amsfonts}
\usepackage{epsfig,rotating}

\def\simeq{
\mathrel{\raise.3ex\hbox{$\sim$}\mkern-14mu\lower0.4ex\hbox{$-$}}
}

\def\ltsima{$\; \buildrel < \over \sim \;$}
\def\simlt{\lower.5ex\hbox{\ltsima}}
\def\gtsima{$\; \buildrel > \over \sim \;$}
\def\simgt{\lower.5ex\hbox{\gtsima}}

\def\msun{{\rm M_{\odot}}}

\def\be{\begin{equation}}
\def\ee{\end{equation}}

\def\del#1{{}}
\def\ltsima{$\; \buildrel < \over \sim \;$}
\def\simlt{\lower.5ex\hbox{\ltsima}}
\def\gtsima{$\; \buildrel > \over \sim \;$}
\def\simgt{\lower.5ex\hbox{\gtsima}}

\newcommand{\apj}{ApJ}
\newcommand{\apjs}{ApJS}
\newcommand{\mnras}{MNRAS}
\newcommand{\aap}{A\&A}
\newcommand{\araa}{ARA\&A}
\newcommand{\apjl}{ApJL}
\newcommand{\aj}{AJ}
\newcommand{\nat}{Nature}

\newcommand{\nar}{New Astron. Rev.}

\title[TDE-powered AGN in dwarf galaxies]{Tidal disruption events can power the observed AGN in dwarf galaxies}
\author[Kastytis Zubovas]{Kastytis Zubovas$^{1,2,\star}$ \\
  $^{1}$Center for Physical Sciences and Technology, Saul\.{e}tekio av. 3, Vilnius LT-10257, Lithuania \\
  $^{2}$Astronomical Observatory, Vilnius University, Saul\.{e}tekio av. 3, Vilnius LT-10257, Lithuania\\
  $^{\star}$ {E-mail:~} {\rm kastytis.zubovas@ftmc.lt} }

\begin{document}

\maketitle

\begin{abstract}

  In recent years, numerous active galactic nuclei have been
  discovered in ever smaller galaxies, questioning the paradigm that
  dwarf galaxies do not harbour central massive black holes. Even if
  such black holes exist, feeding them by gas streams is difficult,
  since star formation should be more efficient than AGN feeding in
  dwarf galaxies. In this paper, I investigate the possibility that
  tidal disruptions of stars are responsible for the observed AGN in
  dwarf galaxies. I show that the expected duty cycles of TDE-powered
  AGN, $f_{\rm AGN} \simgt 0.5\%$, are consistent with observed AGN
  fractions assuming that the occupation fraction in dwarf galaxies is
  close to unity. Furthermore, I calculate the properties of outflows
  driven by TDE-powered AGN under idealised conditions and find that
  they might have noticeable effects on the host galaxies. Outflows
  themselves might not be detectable, except in gas-poor galaxies,
  where they can accelerate to $v_{\rm out} > 100$~km/s, but increased
  gas turbulence, more diffuse density profile and lower star
  formation efficiency can be discovered and used to constrain the
  black hole occupancy fraction and more nuanced effects on dwarf
  galaxy evolution. If massive black holes form from seeds that are
  much more massive than stellar black holes, then their outflows
  should be easily detectable; this result, aided by observations of
  high-redshift dwarf galaxies, provides a potential way of
  determining seed masses of black holes.
  
\end{abstract}


\section{Introduction}

It is now generally well accepted that supermassive black holes
(SMBHs) exist at the centres of all massive galaxies
\citep{Volonteri2012Sci, Graham2016ASSL}. Accretion energy released by
SMBHs during the active galactic nucleus (AGN) phase, in the form of
radiation, winds and jets, can have a significant impact on the
morphology of the surrounding gas and star formation rate of the host
galaxy \citep{Shankar2006ApJ, McNamara2007ARA&A, Zubovas2012ApJ,
  Zubovas2013MNRASb}, making them important elements of galaxy
evolution. Recently, AGN have been detected in ever smaller galaxies
\citep{Greene2006NewAR, Pardo2016ApJ, Mezcua2016ApJ, Mezcua2017IJMPD,
  Mezcua2018MNRAS}, shrinking the mass gap between stellar and
supermassive black holes. These discoveries suggest that AGN may have
significant effects on dwarf galaxy evolution \citep[see
  also][]{Silk2017ApJ} and provide potential ways of determining the
occupation fraction and duty cycle of massive black holes in the
low-mass range.

At first glance, it is difficult to see how SMBHs can be fed
efficiently in dwarf galaxies. The short dynamical times there lead to
rapid gas consumption by star formation and stellar feedback, starving
the SMBH of fuel \citep{Nayakshin2009MNRAS}. This argument explains
the different scalings of central massive objects observed in galaxies
with low and high central velocity dispersions
\citep{Ferrarese2006ApJ, Martin2018arXiv} and has been typically used
as an explanation for the lack of SMBHs in dwarf galaxies. While some
periods of activity are plausible simply due to stochastic variations
in gas orbits, which can lead to gas streams occasionally feeding the
SMBH efficiently, the duty cycle should be low enough that no
significant SMBH growth should be expected in dwarf galaxies over a
Hubble time. This conclusion is supported by recent results of
\citet{Chilingarian2018arXiv, Bellovary2018arXiv}, who find no
correlation between the mass of 305 candidate intermediate mass black
holes and host galaxy stellar masses.

On the other hand, SMBHs can potentially be fed by stellar tidal
disruption events (TDEs). Each disruption event can easily feed the
black hole in a dwarf galaxy at its Eddington rate for decades
\citep{Rees1988Natur, DeColle2012ApJ}, and their cumulative effect may
lead to noticeable growth. Previous research on the topic
\citep{Magorrian1999MNRAS, Wang2004ApJ, Stone2016MNRAS} generally
found that the smallest galaxies with SMBHs should have the highest
rates of TDEs, because smaller galaxies are more compact and have
shorter relaxation times \citep[but see][]{Brockamp2011MNRAS}. The
total mass added to the SMBH by tidal disruption and direct swallowing
of stars can reach $\sim 10^6 \msun$ over a Hubble time
\citep{Magorrian1999MNRAS}, which can be a significant fraction of the
SMBH mass at $z = 0$. The energy released during these disruption
events can, in principle at least, drive outflows throughout the host
galaxy.

In this paper, I investigate the importance of TDEs to the growth of
SMBHs in dwarf galaxies. Using the TDE rates and their dependence on
SMBH mass as derived by \citet{Stone2016MNRAS}, I find that SMBHs with
seed mass $M_{\rm BH}/\msun < 10^5$ can grow by $>10\%$ over the
Hubble time and are kept in an active state for $0.2-2.5\%$ of the
Hubble time by fuelling via TDEs only. This is consistent with
observational results of AGN fraction in dwarf galaxies, assuming that
most dwarf galaxies harbour central massive black holes. The energy
released during these activity episodes can be several orders of
magnitude higher than the gas binding energy of a gas-rich dwarf
galaxy. Therefore, even wind-driven AGN outflows, with an energy
coupling efficiency of $\eta \sim 5\%$, can have a noticeable effect
on the host dwarf galaxy, even though the outflows themselves might be
difficult to detect. Detection of dwarf galaxies perturbed by past AGN
activity, combined with these results, would help determine the
occupation fraction of massive black holes in dwarf galaxies and
potentially help distinguish between massive black hole origin
scenarios.

The paper is structured as follows. In Section \ref{sec:feed}, I
review the argument as to why AGN feeding by gas flows should be
inefficient in dwarf galaxies. In Section \ref{sec:tdes}, I derive the
expected mass growth and energy release caused by TDEs in SMBHs with
different initial masses. In Section \ref{sec:outflows}, I calculate
the details of AGN outflow propagation and show that the impact on the
host galaxy should be noticeable. In Section \ref{sec:discuss}, I
discuss the implications of these results. Finally, I summarize the
results and conclude in Section \ref{sec:concl}.

\section{Feeding AGN in dwarf galaxies with gas streams} \label{sec:feed}

Typically, nuclear activity is instigated by gas streams resulting
from tidal disruption of molecular clouds falling close to the SMBH. A
fraction of the cloud mass is captured by the SMBH gravitational
potential and forms an accretion disc, which can feed the SMBH for
periods of $10^4-10^5$~yr \citep{King2015MNRAS,
  Schawinski2015MNRAS}. However, the infalling gas can also form
stars, and gas consumption by star formation and stellar feedback
competes against SMBH feeding and feedback. \citet{Nayakshin2009MNRAS}
show, using analytical estimates based on relationships between
luminosity, radius and velocity dispersion for large galaxies, that in
galaxies with central velocity dispersion $\sigma_* \lesssim
150$~km~s$^{-1}$, stellar feedback dominates over AGN feedback and the
growth of the nuclear stellar cluster is more efficient than the
growth of the SMBH. This argument is supported by results of numerical
simulations \citep{Finlator2008MNRAS, Schaye2010MNRAS,
  Habouzit2017MNRAS} and observations of a transition in the black
hole mass - galaxy velocity dispersion relationship
\citep{Martin2018arXiv}.

\begin{figure}
  \centering
    \includegraphics[trim = 0 0 0 0, clip, width=0.45\textwidth]{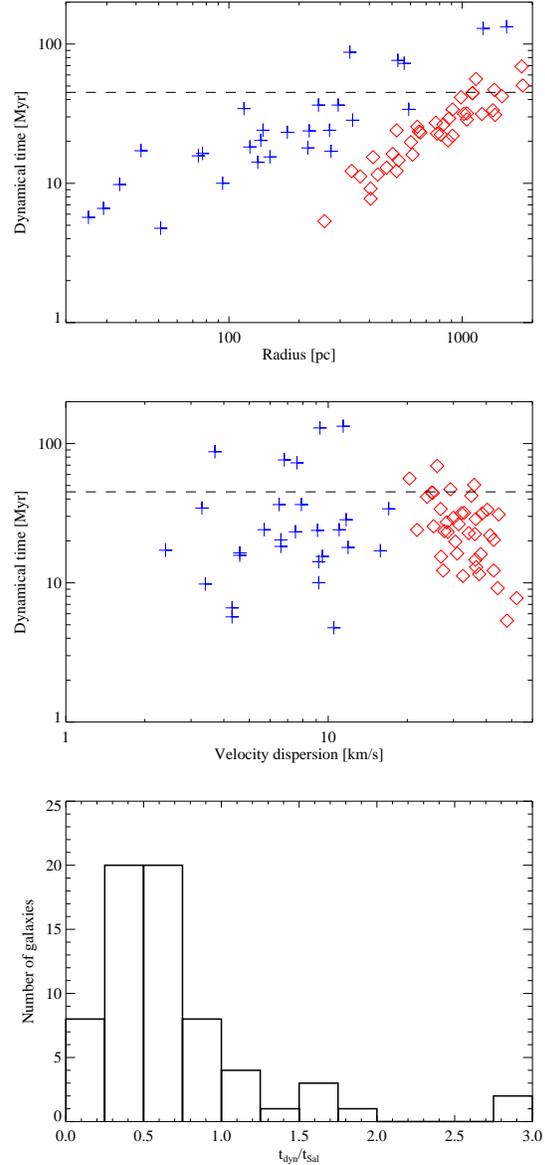}
  \caption{{\em Top}: Dynamical time against effective radius for
    dwarf galaxies in the Local group \citep[blue crosses;
    ][]{Walker2009ApJ} and early-type dwarf galaxies in the Virgo
    cluster \citep[red diamonds; ][]{Toloba2014ApJS}. Dashed
    horizontal line shows $t_{\rm dyn} = t_{\rm Sal} = 45$~Myr. {\em
      Middle}: Dynamical time against velocity dispersion for the same
    galaxies. {\em Bottom}: Histogram of the ratio $t_{\rm dyn}/t_{\rm
      Sal} = t_{\rm dyn}/(45 {\rm Myr})$ for the same galaxies.}
  \label{fig:dgtdyn}
\end{figure}

The same analytical argument can be applied to dwarf
galaxies. Although they follow the baryonic Tully-Fisher relation
$M_{\rm b} \propto v^4$ \citep{Iorio2017MNRAS}, with $M_{\rm b}$ the
total baryon mass and $v$ the circular velocity, the velocity
dispersions exhibit a much larger scatter than typical for large
galaxies \citep[e.g.,][]{Kourkchi2012MNRAS}. Kinematical data of Local
group dwarf spheroidals \citep[Table 1 and references therein
  in][]{Walker2009ApJ} shows only mild correlation between half-mass
radius $r_{\rm half}$ and velocity dispersion $\sigma$, and
essentially no correlation between dynamical time $t_{\rm dyn} =
r_{\rm half}/\sigma$ and $\sigma$. However, only five of the 28
galaxies presented in that data set have $t_{\rm dyn} > t_{\rm Sal}$,
where $t_{\rm Sal} \simeq 45$~Myr is the Salpeter time. Similarly,
only 4-6 out of 39 dwarf early-type galaxies in the Virgo cluster
\citep{Toloba2014ApJS} have $t_{\rm dyn} > t_{\rm Sal}$. In Figure
\ref{fig:dgtdyn}, I show the relationship between $t_{\rm dyn}$ and
effective or half-mass radius (top panel) and velocity dispersion
(middle panel), as well as a histogram of the ratio $t_{\rm
  dyn}/t_{\rm Sal}$ (bottom panel), for these two dwarf galaxy
samples. $84\%$ of dwarf galaxies have dynamical times shorter than
$t_{\rm Sal}$, and $73\%$ have $t_{\rm dyn} < 0.75 t_{\rm
  Sal}$. Galaxies with $t_{\rm dyn} > t_{\rm Sal}$ are the largest in
both samples, and it is certainly possible that dynamical times in
their central regions, inside the sphere of influence of any putative
central massive black hole, are significantly shorter.

Such approximate estimates suggest that SMBH feeding is less efficient
than star formation in most dwarf galaxies. Gas falling in toward the
centre of a dwarf galaxy is predominantly consumed by star formation,
and stellar feedback disperses the gas more efficiently than AGN
feedback. This conclusion is supported by observations of the
correlation between star formation rate (SFR) and black hole accretion
rate (BHAR). The ratio BHAR/SFR decreases with decreasing stellar
mass, to $\langle BHAR \rangle / \langle SFR \rangle < 10^{-4}$ at
$M_* \simlt 10^{9.7} \msun$ \citep{Yang2018MNRAS} and potentially
$\langle BHAR \rangle / \langle SFR \rangle < 10^{-4.5}$ at $M_*
\simlt 10^{8.2} \msun$ \citep{Yang2017ApJ}. The power of stellar wind
feedback \citep{Leitherer1992ApJ} is equal to AGN wind feedback power
\citep{King2010MNRASa} if $\langle BHAR \rangle / \langle SFR \rangle
\simeq 3.5 \times 10^{-4}$ (Zubovas 2018, accepted), therefore
low-mass galaxies should indeed be affected more strongly by stellar
feedback than AGN feedback, at least on average. Another evidence in
favour of this conclusion comes from numerical simulations of galaxy
evolution, which show that black holes in dwarf galaxies are starved
of gas \citep{Bonoli2016MNRAS, Trebitsch2017arXiv,
  Bellovary2018arXiv}.

Stellar winds can in principle also feed AGN
\citep[e.g.,][]{Davies2012JPhCS}. If there is a star formation episode
near the galactic nucleus, massive star winds may feed the AGN for
several Myr afterward \citep{Coker1997ApJ, Cuadra2005MNRAS,
  Davies2007ApJ, Cuadra2008MNRAS}. The feeding rate, however, is
unlikely to be very large: \citet{Cuadra2005MNRAS, Cuadra2008MNRAS}
find an expected rate of a few times $10^{-6} \; \msun$~yr$^{-1}$ for
the conditions of the centre of the Milky Way. Assuming a radiative
efficiency of $10\%$, this accretion rate corresponds to a bolometric
luminosity of $L_{\rm bol} \sim 10^{40}$~erg~s$^{-1}$, close to the
lower limit of detection for AGN in dwarf galaxies
\citep[cf.][assuming a bolometric correction $B =
  10$]{Mezcua2016ApJ,Mezcua2018MNRAS}. In a dwarf galaxy, with a
smaller nuclear star cluster and a smaller central black hole, the
stellar wind accretion rate is likely to be much lower than the result
of \citet{Cuadra2005MNRAS}.

Winds of low-mass main sequence or red giant stars can also feed the
black hole \citep[e.g.,][]{Shull1983ApJ}. The mass loss rate of a
single red giant star can reach up to $10^{-4} \; \msun$~yr$^{-1}$
\citep{Willson2000ARA&A}. The wind capture rate by the SMBH is highly
uncertain, however; as an upper limit, we may consider the same
simulations by \citet{Cuadra2008MNRAS}, where the black hole accretion
rate is $\simlt 0.3\%$ of the total mass injection rate by winds from
massive stars well within the central parsec. The fraction drops by
another factor $\sim 2$ when we consider that feedback outbursts can
shut off accretion for extended periods of time
\citep{Cuadra2015MNRAS}. We can therefore estimate that a single red
giant star in the star cluster around the SMBH feeds it at an average
rate $\dot{M}_{\rm RG} < 1.5\times 10^{-7} \; \msun$~yr$^{-1}$. In
order to feed a $10^5 \; \msun$ black hole at $\dot{M} = 0.01
\dot{M}_{\rm Edd} = 2.2 \times 10^{-5} \; \msun$~yr$^{-1}$, there
should be more than a hundred red giants in the surrounding nuclear
star cluster, all emitting winds close to the peak rate. Since this
peak mass loss rate cannot last for much longer than a few thousand
years per star, this situation appears unlikely. As a more
conservative estimate, we can consider the initial-final mass
relations for stars of various masses and calculate that a stellar
population loses $\sim 1/3$ of its initial mass via outflows from
stars with $M_{\rm init} < 8 \; \msun$ \citep{Cummings2018ApJ}.
Considering a similar capture fraction as before, we see that only
$\sim 10^{-3}$ of the nuclear star cluster mass will be added to the
black hole mass over the cluster's lifetime. This fraction is
negligible for any reasonable BH/cluster mass ratio. I therefore
conclude that stellar winds are unlikely to contribute significantly
to the growth of central black holes, although sporadic activity
episodes may be generate by accretion of clumpy wind material.

\section{Tidal disruption around IMBHs} \label{sec:tdes}

\subsection{TDEs as a power source for AGN}

Even though, as shown in the previous section, AGN feeding by gas
streams appears to be inefficient in dwarf galaxies, there is evidence
that AGN feedback can affect the dwarf hosts \citep{Penny2018MNRAS}.
AGN feedback can potentially explain a large number of cosmological
problems related to dwarf galaxies \citep{Silk2017ApJ}. Therefore, it
is interesting to consider tidal disruption events (TDEs) as a
possible source of AGN fuel.

Typically, TDEs are considered only in terms of the flares with
characteristic light curves, rather than significant contributors to
the SMBH growth. However, the fallback rate of the disrupted material
increases with decreasing SMBH mass, and exceeds the Eddington limit
at $M_{\rm BH} \leq 2.4 \times 10^7 \; \msun$ \citep{DeColle2012ApJ}.
At lower masses, the phase of significant black hole feeding rate
becomes ever longer. In particular, each TDE results in
super-Eddington accretion for a time \citep{Stone2016MNRAS}
\begin{equation} \label{eq:tedd}
  t_{\rm Edd} = 5.1 \eta_{0.1}^{0.6} M_5^{-0.4} m_*^{0.2} r_*^{0.6} {\rm yr},
\end{equation}
where $\eta \equiv 0.1\eta_{0.1}$ is the radiative efficiency of
accretion, $M_5 \equiv M_{\rm BH}/10^5 \; \msun$ is the BH mass, and
$m_*$ and $r_*$ are the mass and radius of the disrupted star in Solar
units.

During the super-Eddington fallback phase of the TDE, the flare
luminosity is not directly proportional to the mass flow rate, because
material builds up an accretion disc, which is then depleted on a
viscous timescale, potentially powering an observable flare for $t >
100$~yr \citep{RamirezRuiz2009ApJ, Hayasaki2013MNRAS}. During the
super-Eddington accretion phase, the black hole behaves as a ULX, with
luminosity varying as $L = L_{\rm Edd}\left(1 + {\rm
  ln}\left(\dot{M}/\dot{M_{\rm Edd}}\right)\right)$
\citep{Shakura1973A&A, King2016MNRASb}. This results in a luminosity
evolution $L \propto t^{-\alpha}$, where $\alpha$ varies between $\sim
1/4$ initially to the canonical value $5/3$ at late times, when the
accretion rate drops below Eddington; however, \citet{Piran2015ApJ}
argue that the flare is powered by energy release during the build-up
of the accretion disc, which would lead to $L \propto t^{-5/3}$ at all
times. In any case, the accretion disc can be expected to cool
efficiently as long as $L > 10^{-2} L_{\rm Edd}$
\citep[e.g.,][]{Heckman2014ARA&A}, forming a geometrically thin disc
and driving a wide-angle wind \citep{King2003MNRASb,
  King2016MNRASb}. This wind does not affect the disc feeding rate,
since material falls on to the disc predominantly in the disc plane,
but may affect the larger scale environment of the black hole.

The total energy released during a single TDE is uncertain. In the
simplest picture, if approximately half of a Solar-mass star is
eventually accreted with a $10\%$ radiative efficiency, the expected
energy release is of order $10^{53}$~erg; however, most observed TDEs
have total energies 1-2 orders of magnitude lower
\citep[e.g.,][]{Gezari2012Natur, Holoien2016MNRAS,
  Eftekhari2018ApJ}. This discrepancy may be explained if most of the
energy is emitted in extreme UV (EUV), which is difficult to observe,
or in jets beamed away from Earth \citep{Lu2018arXiv}. Alternatively,
the disc wind may carry away a significant amount of mass and the
total accreted mass may be only a small fraction of the initial mass
of the disrupted star. In the following, I take the expression for the
total radiated energy from \cite{Stone2016MNRAS}:
\begin{equation}
  E_{\rm TDE} = 2\times10^{51} \eta_{0.1}^{0.6} M_5^{0.6} m_*^{0.2} r_*^{0.6} {\rm erg},
\end{equation}
which corresponds to an accreted mass
\begin{equation} \label{eq:deltam}
  \Delta M_{\rm BH} = \frac{E_{\rm TDE}}{\eta c^2} = 1.1 \times 10^{-2} \eta_{0.1}^{-0.4} M_5^{0.6} m_*^{0.2} r_*^{0.6} \msun.
\end{equation}
This is a conservative estimate, where only $\sim 1\%$ of the mass of
the star ends up powering the TDE flare. In fact, the radiated energy
estimate is $E_{\rm TDE} \simeq L_{\rm Edd} t_{\rm Edd}$, but it may
be significantly higher if we account for the buildup of the accretion
disc during the super-Eddington feeding phase \citep{Lin2017NatAs}.
Also, including the potential EUV emission increases the total energy,
and hence the effect on the galaxy. In order to reduce possible
uncertainties, I also do not account for the possibility of a phase of
hyper-Eddington accretion, which may be relevant for black holes with
$M \lesssim 8\times10^4 \msun$, which experience $\dot{M}_{\rm peak} >
5000 \dot{M}_{\rm Edd}$ \citep{Inayoshi2016MNRAS}.

The goal of this paper is to provide a proof of concept that
TDE-powered AGN can account for a significant part of the observed AGN
population in dwarf galaxies, and that outflows caused by such AGN can
have noticeable effects on the host galaxies. Therefore, I do not
consider the precise details of individual TDEs, instead focusing on
the conservative estimates for their numbers, durations and energy
release.

\subsection{TDE rates and properties in dwarf galaxies} \label{sec:tde_rates}

There are significant uncertainties regarding the TDE rate and its
dependence on host galaxy properties. It is generally expected that
the TDE rate should be higher for lower-mass black holes
\citep{Wang2004ApJ, Stone2016MNRAS}; however, the scaling with BH mass
may be opposite in star clusters \citep{Sakurai2018arXiv}. In
particular, for cored galaxies, the expected TDE rate is
\citep{Stone2016MNRAS}
\begin{equation} \label{eq:ntde}
  \dot{N}_{\rm TDE} = 6.6\times 10^{-5} M_5^{-0.247} {\rm yr}^{-1}.
\end{equation}

The observed TDE rate is generally lower, of order $N_{\rm TDE, obs}
\sim 10^{-5} {\rm yr}^{-1}$ per galaxy. \citet{Stone2016MNRAS}
provides an extensive discussion regarding the possible sources of
this discrepancy. \citet{Mageshwaran2018arXiv} provide a potential
solution to the discrepancy, claiming that averaging the theoretically
predicted results over the black hole mass function, and considering
the detection probability of each event, leads to agreement between
predicted and observed rates. In particular, the predicted rate,
averaged over the galaxy population, can be a factor 2 higher than
observed. Therefore, I will keep the estimate of eq. (\ref{eq:ntde})
for the rest of the calculations, and discuss the implications of it
being lower in the Discussion section.

\begin{figure}
  \centering
    \includegraphics[trim = 0 0 0 0, clip, width=0.45\textwidth]{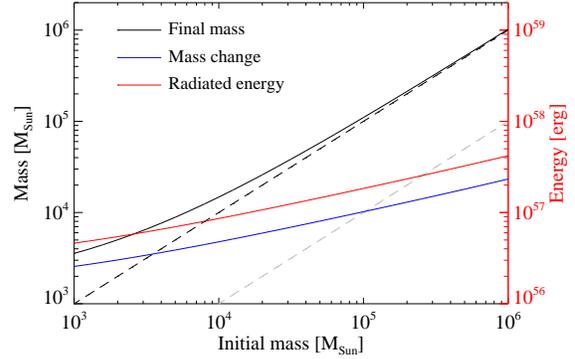}
  \caption{Expected growth of black holes over Hubble time via tidal
    disruption of stars. Black line show the initial-final mass
    relation, blue line shows the change in mass, red line shows the
    total radiated energy (scale on the right). Dashed black line
    shows a 1-to-1 correspondence between initial mass and final mass
    or mass change, grey dashed line shows 10\% of the initial mass.}
  \label{fig:massgrowth}
\end{figure}

Combining equations (\ref{eq:ntde}) and (\ref{eq:deltam}) leads to an
average mass growth rate over timescales much longer than
$\dot{N}_{\rm TDE}^{-1}$:
\begin{equation} \label{eq:dmdt}
  \dot{M}_{\rm BH} = \dot{N}_{\rm TDE} \Delta M_{\rm BH} \simeq 7.3 \times 10^{-7} \eta_{0.1}^{-0.4} M_5^{0.353} m_*^{0.2} r_*^{0.6} \msun {\rm yr}^{-1}.
\end{equation}
Although mass growth via TDEs is stochastic and extremely bursty, the
above expression can be used to estimate the total mass growth of the
SMBH over the Hubble time. Rearranging the equation and integrating
leads to an initial-final mass relation:
\begin{equation} \label{eq:m0mf}
  M_{\rm f,5} = \left(M_{\rm 0,5}^{0.647} + 4.7 \times 10^{-3} \eta_{0.1}^{-0.4} m_*^{0.2} r_*^{0.6} \frac{t}{{\rm Gyr}}\right)^{1.546},
\end{equation}
where $M_{\rm 0,5}$ and $M_{\rm f,5}$ are the initial and final masses
in units of $10^5 \msun$. Figure \ref{fig:massgrowth} shows this
relation for initial masses $10^3 < M_0/\msun < 10^6$, as well as the
difference between the two masses, assuming $\eta_{0.1} = m_* = r_* =
1$. As expected, the total mass gain is not very large, but can be
significant for the smallest seed black holes. In particular, an $M_0
= 10^4 \msun$ black hole grows by a factor $M_{\rm f}/M_0 \simeq
1.48$, while an $M_0 = 10^5 \msun$ one grows by only $\sim 10\%$.

The total energy released during TDEs over the Hubble time is directly
proportional to the mass change, i.e.
\begin{equation} \label{eq:erad}
  E_{\rm rad} = \eta \Delta M c^2 = 1.8 \times 10^{53} \eta_{0.1} {\rm erg} \frac{\Delta M}{\msun}.
\end{equation}
This value is plotted as a red line in Figure \ref{fig:massgrowth},
with scale on the right. We see that TDEs can result in release of
$5\times 10^{56} - 4\times 10^{57}$~erg over the Hubble time. In the
next section, I show that this energy release can have noticeable
effect on the gas distribution in the host galaxy.

The duration of the TDE flare, in particular the super-Eddington phase
(eq. \ref{eq:tedd}), allows us to calculate the expected duty cycle of
the AGN and the expected fraction of time for which the host galaxy is
observed as being active. Using the standard TDE accretion rate
relation $\dot{M} \propto t^{-5/3}$ \citep{Rees1988Natur}, which is
applicable at late times at least \citep{Lodato2009MNRAS}, we have
that the duration for which the AGN is brighter than some threshold
$L_{\rm thr} < L_{\rm Edd}$ is
\begin{equation}\label{eq:tthr}
  t_{\rm thr} = t_{\rm Edd} \left(\frac{L_{\rm Edd}}{L_{\rm thr}}\right)^{3/5}.
\end{equation}
Using this equation and the typical AGN threshold $L_{\rm AGN} > 0.01
L_{\rm Edd}$ \citep[e.g.,][]{Heckman2014ARA&A} gives
\begin{equation}\label{eq:tagn}
  t_{\rm AGN} \simeq 15.8 t_{\rm Edd} \simeq 80.8 \eta_{0.1}^{0.6} M_5^{-0.4} m_*^{0.2} r_*^{0.6} {\rm yr}
\end{equation}
per TDE. The duty cycle is then
\begin{equation} \label{eq:fagn}
  f_{\rm AGN} = \dot{N}_{\rm TDE} t_{\rm AGN} \simeq  5.3\times10^{-3} \eta_{0.1}^{0.6} M_5^{-0.647} m_*^{0.2} r_*^{0.6},
\end{equation}
a value rather similar to the currently observed fraction of dwarf
galaxies hosting AGN \citep[$0.4-3\%$, cf.][]{Pardo2016ApJ,
  Mezcua2018MNRAS}. If, instead, a luminosity threshold is adopted,
e.g. $L_{\rm X} > 10^{41}$~erg~s$^{-1}$ \citep{Pardo2016ApJ}, with a
bolometric correction $L_{\rm bol} \equiv 10 B_{10} L_{\rm X}$, we have
\begin{equation} \label{eq:fagnx}
  \begin{split}
    f_{\rm AGN,X} &= \dot{N}_{\rm TDE} t_{\rm Edd} \left(\frac{L_{\rm Edd}}{B L_{\rm X}}\right)^{3/5} \\
    & \simeq 1.6\times10^{-3} \eta_{0.1}^{0.6} M_5^{-0.047} m_*^{0.2} r_*^{0.6} B_{10}^{-0.6}.
  \end{split}
\end{equation}
This value is somewhat smaller than current observational
estimates. However, the threshold is somewhat subjective: using
$L_{\rm X} > 10^{39}$~erg~s$^{-1}$ \citep[following][]{Mezcua2016ApJ,
  Mezcua2018MNRAS}, I find $f_{\rm AGN,X} = 0.025 \eta_{0.1}^{0.6}
M_5^{-0.047} m_*^{0.2} r_*^{0.6} B_{10}^{-0.6}$, consistent with
observed numbers.

Four effects can increase the expected duty cycle. First of all, more
detailed modelling of fallback of material, including its
self-gravity, shows that the disrupted gas stream can become clumpy
and provide a higher peak fallback rate \citep{Coughlin2015ApJ,
  Coughlin2016MNRAS}. Therefore, a larger fraction of TDEs might have
prompt emission rising above the detection threshold. Furthermore, the
super-Eddington fallback rate can persist for longer, by a factor
$\sim 1.5$, than given by simple analytical estimates
\citep[eq. \ref{eq:tedd}; cf.][]{Wu2018MNRAS}. Thirdly, if we assume
that the falling material circularizes into a disc which then feeds
the black hole via viscous transport, the duration of activity can
extend even further, since the viscous timescale of a $10^5 \msun$
black hole is a few decades \citep{RamirezRuiz2009ApJ}. Such behaviour
has been observed in super-Eddington TDEs \citep{Lin2017NatAs},
although currently the numbers are small and statistical analysis of
their frequency and effects is difficult. Finally, the flaring rate
may be somewhat higher if disruption of red giants is included.

\subsection{Observable properties of TDE-powered AGN} \label{sec:obs_properties}

These results suggest that TDE-powered AGN episodes may account for a
significant fraction of all observed AGN in dwarf galaxies. As larger
samples and longer-term observations of AGN in dwarf galaxies become
available, it should be possible to distinguish TDE-powered episodes
from other kinds of nuclear activity. This will allow for the testing
of this prediction. In particular, the TDE-powered AGN flares should
have certain properties that distinguish them from feeding via stellar
winds or interstellar gas streams:

\begin{itemize}
\item The mass of the disrupted star is of order $M \sim 1 \; \msun$,
  while other kinds of gas streams do not have such a limitation;
  therefore, the mass of a TDE-fed accretion disc is always small,
  much smaller than the self-gravity limit $M_{\rm d} \lesssim H/R
  M_{\rm BH} \sim 10^{-3} M_{\rm BH}$. This affects the AGN spectrum,
  since the luminosity of the disc, especially its outer regions, is
  likely to be smaller than in the case of a disc fed by gas streams.

\item The disc feeding rate, and therefore also the disc mass, change
  significantly on timescales of years and decades. Even though at
  early times, neither the feeding rate nor the flare luminosity
  evolve as the classical estimate $\dot{M} \propto t^{-5/3}$
  predicts, at late times this becomes true. Therefore, observations
  spanning multiple years and multiple dwarf AGN should find some AGN
  in the decay phase of their lightcurve. The upper limit to the
  fraction of AGN in the fading phase is
  \begin{equation}
    f_{\rm fade} < \frac{t_{\rm AGN} - t_{\rm Edd}}{t_{\rm AGN}} \simeq 0.94,
  \end{equation}
  where $t_{\rm AGN}$ and $t_{\rm Edd}$ are taken from equations
  (\ref{eq:tagn}) and (\ref{eq:tedd}), respectively. Accounting for
  the super-Eddington disc feeding phase decreases this fraction,
  since the `plateau' phase of the flare is extended by the viscous
  timescale of the accretion disc \citep{Lin2017NatAs}.

\item In addition to changes on decade timescales, flares should show
  significant spectral variability on timescales much shorter than AGN
  fed by gas streams \citep{Lin2017NatAs}.

\item Flares reach their peak luminosity very rapidly, on the orbital
  timescale of the bound disrupted material, therefore AGN in dwarf
  galaxies appearing abruptly would be better explained as TDEs rather
  than as being fed by gas streams.
\end{itemize}

While none of these properties can by itself identify TDE-powered AGN
from those powered by gas streams, multiple lines of evidence may be
used to distinguish between the two modes. If the TDE-powered AGN
fraction among all dwarf galaxies turns out to be smaller than the
estimates in equations (\ref{eq:fagn}) and (\ref{eq:fagnx}), these
results would also give another estimate of the massive black hole
occupation fraction in dwarf galaxies. As the amount of data increases
and our understanding of TDE properties grows, it should eventually be
possible to determine the massive black hole occupation fraction in
subsets of dwarf galaxies selected by mass or other properties. This
would help significantly improve our understanding of the formation
and growth of SMBHs throughout the Universe.

\section{AGN outflows in dwarf galaxies} \label{sec:outflows}

Independently of how an AGN is powered, its luminosity can have a
significant effect on the host galaxy. If accretion proceeds via a
thin disc, a disc wind can reach quasi-relativistic velocities $v_{\rm
  w} \sim 0.1 c$ and drive large-scale outflows which remove gas from
the host galaxy \citep{Zubovas2012ApJ} and potentially trigger
starbursts \citep{Zubovas2013MNRASb}. In this section, I investigate
the expansion of spherically-symmetric wind driven AGN outflows in
dwarf galaxies using a 1D code and show that even AGN powered purely
by tidal disruption of stars may have significant effects on their
host galaxies. In TDEs, jets may have a similar energy output to disc
radiation, but I do not include them in these calculations. I discuss
the potential importance of jet feedback in Section
\ref{sec:other_modes}.

\subsection{Physical model} \label{sec:physmodel}

The wind outflow model is described in detail in
\citet{King2010MNRASa} and \citet{Zubovas2012ApJ}. It is based on the
observationally established fact that AGN accreting via a thin disc
drive quasi-relativistic winds, which have velocities of order $v_{\rm
  w} \sim 0.1 c$ and mass flow rates $\dot{M}_{\rm w}$ comparable to
the black hole accretion rate $\dot{M}_{\rm acc}$
\citep{Pounds2003MNRASa, Tombesi2010A&A, Tombesi2010ApJ}. Within the
model, the wind self-regulates to keep an optical depth $\sim 1$,
which means that the photons of the AGN radiation field scatter on
average once before escaping, and therefore
\begin{equation}
\dot{M}_{\rm w} v_{\rm w} = \frac{L_{\rm AGN}}{c} = \eta \dot{M}_{\rm acc} c.
\end{equation}
As the wind reaches the interstellar medium (ISM), a two-shock system
develops, with a forward shock driven into the ISM and a backward
shock driven into the wind. If the backward shock cools efficiently,
most likely via inverse-Compton scattering, most of the kinetic energy
of the wind, $\dot{E}_{\rm w} \simeq \eta L_{\rm AGN}/2$, is
radiated away and a momentum-driven outflow develops. Alternatively,
if the wind doesn't cool, most of the energy is transferred to the ISM
and an energy-driven outflow is formed. Cooling of the shocked wind is
efficient only very close to the AGN \citep[perhaps as close as
  $R_{\rm C} < 1$~pc, cf.][]{Faucher2012MNRASb}, so the galaxy at large is
affected by energy-driven outflows.

Assuming that the outflow is perfectly adiabatic and expanding in a
spherically symmetric background potential, pushing a spherically
symmetric gas distribution, the equation of motion is \citep[for
  derivation see][]{Zubovas2016MNRASb}:
\begin{equation} \label{eq:eom}
  \begin{split}
    \dddot{R} &= \frac{\eta L_{\rm AGN}}{M R} - \frac{2\dot{M} \ddot{R}}{M} - \frac{3\dot{M} \dot{R}^2}{M R} - \frac{3\dot{R} \ddot{R}}{R} - \frac{\ddot{M} \dot{R}}{M} \\ & +\frac{G}{R^2}\left[\dot{M} + \dot{M}_{\rm b} + \dot{M}\frac{M_{\rm b}}{M} - \frac{3}{2}\left(2M_{\rm b}+M\right)\frac{\dot{R}}{R}\right].
  \end{split}
\end{equation}
Here, $\dot{M}\equiv \dot{R}\partial M/\partial R$ and $\ddot{M}
\equiv \ddot{R}\partial M/\partial R + \dot{R} ({\rm d}/{\rm
  d}t)\left(\partial M/\partial R\right)$. The first term on the right
hand side is the driving term, the next four terms arise from the
$p{\rm d}V$ work and the kinetic energy of the outflow, and the terms
in square brackets appear due to the force of gravity and work against
the gravitational potential.

The assumption of spherical symmetry is, of course, a gross
simplification of a real dwarf galaxy. A large fraction, perhaps the
majority, of dwarf galaxies have irregular morphologies
\citep{Ann2017JKAS}. The TDE may be accompanied by a jet which, by its
nature, is a non-spherical phenomenon. Nevertheless, a spherically
symmetric numerical model can give important insight into the
processes happening in the system in question. In particular, one can
obtain the typical outflow velocities and average velocities,
pressures, as well as energy and momentum rates, which provide
information on the expected observability of outflows, their escape
from the galaxy and potential impact on star formation in the
host. The expansion of outflows driven by wide-angle AGN winds and
supernovae has been investigated extensively via 1D spherically
symmetric numerical means \citep[e.g.,][]{Sharma2013ApJ,
  Zubovas2016MNRASb, Igarashi2017MNRAS, Dashyan2018MNRAS}. In the
Discussion section, I consider the possible effects of asymmetries in
the gas distribution surrounding the black hole.

\subsection{Numerical implementation} \label{sec:nummodel}

\begin{figure}
  \centering
    \includegraphics[trim = 0 0 0 0, clip, width=0.49\textwidth]{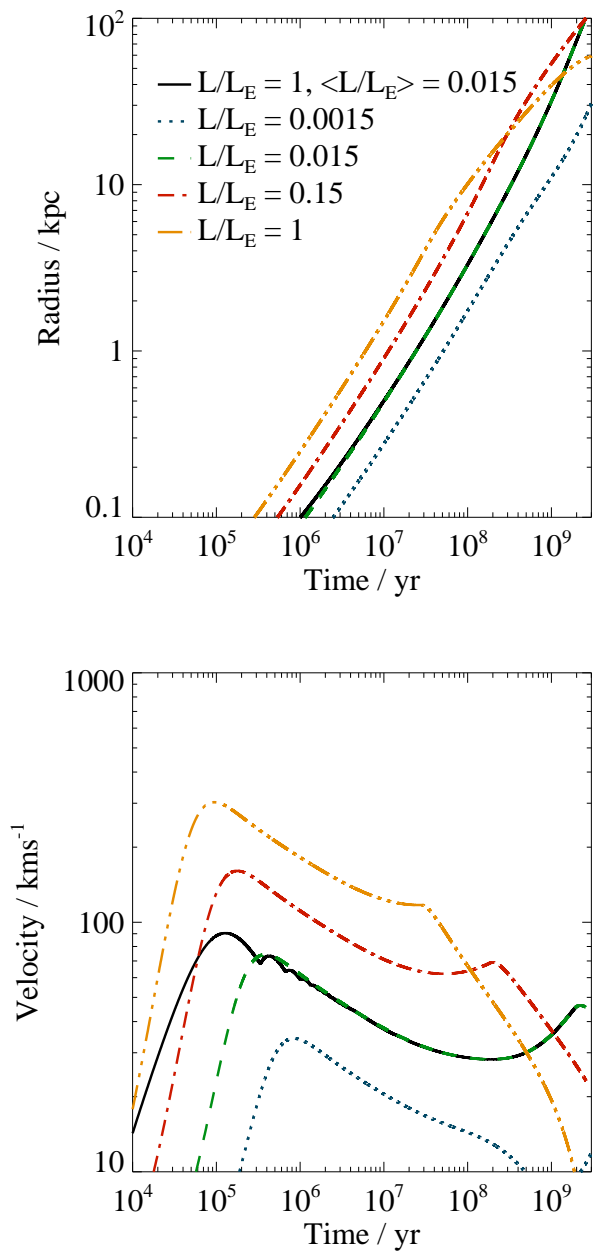}
  \caption{Outflow radius (top) and velocity (bottom) against time in
    simulations with bursty AGN luminosity history (black solid line)
    and several cases of continuous AGN activity, all limited by total
    energy output.}
  \label{fig:twopart_bursts}
\end{figure}

The equation of motion can be integrated numerically to provide all
the salient outflow properties: $R\left(t\right)$, $v\left(t\right)$,
$\dot{M}\left(t\right)$ and so on. I use a third order Taylor
integration scheme with adaptive timesteps chosen according to the
criterion $\Delta t = 0.1 \times {\rm min}\lbrace R/\dot{R},
\dot{R}/\ddot{R}, \ddot{R}/\dddot{R}, t_{\rm q}\rbrace$, where $t_{\rm
  q}$ is the AGN episode duration (see below). This integration
preserves the adiabaticity of the system and recovers analytic
solutions in idealised cases \citep{Zubovas2011MNRAS,
  Zubovas2016MNRASb} and only fails in situations where one or more
parameters of the system change abruptly, but such situations only
arise due to other assumptions in the model, such as perfect spherical
symmetry.

In all the simulations presented here, the galaxy is defined by its
total mass $M_{\rm tot}$ and a gas fraction $f_{\rm g} \equiv M_{\rm
  g}/M_{\rm tot}$. The gas and underlying collisionless mass (dark
matter and stars) is distributed in an NFW profile with parameters
dependent on redshift $z$. Following \citet{Dutton2014MNRAS}, I adopt
a relationship for concentration:
\begin{equation}
c = 10^{a+b{\rm log}\left(M_{12}h\left(z\right)\right)},
\end{equation}
where
\begin{equation}
  \begin{split}
    a &= 0.52+\left(0.905-0.52\right){\rm exp}\left(-0.617z^{1.21}\right); \\
    b &= -0.10+0.026z; \\
    h\left(z\right) &= h_0 \left(\Omega_\Lambda+\Omega_{\rm m}\left(1+z\right)^3\right)^{1/2},
  \end{split}
\end{equation}
and, following \citet{Bryan1998ApJ}, a relation for the virial
overdensity:
\begin{equation}
\Delta_{\rm vir} = 18 \pi^2 + 82x -39x^2,
\end{equation}
where
\begin{equation}
x = 0.308\frac{\left(1+z\right)^3}{h\left(z\right)^2/0.678^2}-1.
\end{equation}
At each timestep, I update the redshift using the
relation
\begin{equation}
  \Delta z = - H_0 \left(1+z\right) \left(\Omega_\Lambda+\Omega_{\rm
    m}\left(1+z\right)^3\right)^{1/2} \Delta t.
\end{equation}
I use the most recent Planck cosmological parameters: $H_0 \equiv 100
h_0 $~km~s$^{-1}$~Mpc$^{-1} = 67.8$~km~s$^{-1}$~Mpc$^{-1}$,
$\Omega_{\rm m} = 0.308$, $\Omega_\Lambda = 0.692$
\citep{Planck2016A&A}.

At the centre of the gas distribution, I put an SMBH with mass $M_{\rm
  BH}$. Nuclear activity is described by peak Eddington ratio $l =
L_{\rm AGN}/L_{\rm Edd}\left(M_{\rm BH}\right)$ and episode duration
$t_{\rm q}$. All simulations begin at $z_{\rm init} = 0.247$ and run
for $t_{\rm tot} = 3$~Gyr, giving the final $z_{\rm fin} = 0$. The
results do not change appreciably if the initial redshift is
increased; I checked this with $z_{\rm init} = 2$ and $z_{\rm init} =
3$. This happens because the evolution of the halo parameters with
redshift is rather weak.

\subsection{Results} \label{sec:outflow_results}

\subsubsection{Bursty and continuous feedback}

At first, I investigate the possible difference between long duration
AGN episodes, such as would be caused by feeding of the SMBH via gas
streams, and bursty feeding through TDEs. I choose a galaxy with total
mass $M_{\rm tot} = 3\times 10^9 \msun$ with $f_{\rm g} = 0.16$,
i.e. $M_{\rm gas} = 4.8\times10^8 \msun$, harbouring a central SMBH
with $M_{\rm BH} = 10^4 \msun$. I run five simulations in total. One
has the BH experiencing Eddington-limited accretion episodes lasting
$t_{\rm AGN} = 5000$~yr every $t_{\rm rep} = 0.33$~Myr, giving an
average Eddington ratio $\langle l\rangle \simeq 0.015$. Although the
values of both $t_{\rm AGN}$ and $t_{\rm rep}$ are much longer than
those derived for TDE-powered AGN episodes, they still produce a large
number of comparatively short AGN episodes, while allowing the
simulation to run in reasonable time without using too much memory for
storing the time evolution of outflow parameters. Four other
simulations each have a single activity episode with $l = \lbrace
0.0015, 0.015, 0.15, 1\rbrace$. AGN is shut off forever once the total
energy released by the AGN reaches $E_{\rm tot} = 1.15\times
10^{57}$~erg. This is equivalent to the AGN injecting a kinetic energy
$E_{\rm kin} = 0.05 E_{\rm tot} = 5.76 \times 10^{55}$~erg into the
gas, which is $1.5$ times the binding energy of the gas in the chosen
potential. The switchoff time for the five simulations is $t_{\rm off}
= \lbrace 1.9, 18.7, 1.9, 0.19, 0.028\rbrace$~Gyr; in the second
simulation, $t_{\rm off} > t_{\rm H}$, so the AGN never switches off.

Figure \ref{fig:twopart_bursts} shows the propagation of outflows in
these simulations. There are noticeable differences among the
simulations with different Eddington ratios: simulations with brighter
AGN produce faster outflows. Brighter AGN also produce more massive
and energetic outflows. On the other hand, the fraction of AGN energy
output converted to outflow kinetic energy is $\sim 2\%$ in all
simulations, independently of the Eddington ratio. At late times,
outflow radii in different simulations become more similar, suggesting
that total injected energy is also an important parameter in
determining the long-term outflow evolution.

The most important result can be seen by comparing the bursty
simulation (black solid line) with the $l = 0.015$ simulation (green
dashed line). Here, we see that burstiness has very little effect on
outflow radius and velocity, except at very early times. Therefore it
is the average Eddington ratio which determines the outflow
properties. Motivated by this result, in subsequent simulations I use
an average AGN luminosity, rather than bursty AGN luminosity history,
to investigate outflows driven by TDE-powered AGN.

\subsubsection{TDE-powered outflows}

\begin{table*}
\begin{tabular}{c | c c c c c c c c}
Model ID & $M_{\rm tot}/\msun$ & $M_{\rm gas}/\msun$ & $M_{\rm BH,0}/\msun$ & $E_{\rm b}$/erg & $E_{\rm SN}$/erg & $E_{\rm AGN}$/erg & $E_{\rm b}/E_{\rm AGN}$ & $E_{\rm SN}/E_{\rm AGN}$ \\
\hline
\hline
M3e8     & $3\times10^8$    & $4.79\times10^7$ & $34$           & $3.3\times 10^{53}$ & $1.2\times 10^{54}$ & $2.8\times 10^{56}$ & $1.2\times10^{-3}$ & $4.3\times10^{-3}$ \\
M1e9     & $10^9$           & $1.59\times10^8$ & $200$          & $2.4\times 10^{54}$ & $7.1\times 10^{54}$ & $3.3\times 10^{56}$ & $7.3\times10^{-3}$ & $0.022$ \\
M3e9     & $3\times10^9$    & $4.76\times10^8$ & $980$          & $1.5\times 10^{55}$ & $3.5\times 10^{55}$ & $4.4\times 10^{56}$ & $0.034$            & $0.080$ \\
M1e10    & $10^{10}$        & $1.58\times10^9$ & $6.2\times10^3$ & $1.1\times 10^{56}$ & $2.2\times 10^{56}$ & $7.2\times 10^{56}$ & $0.15$            & $0.31$ \\
M3e10    & $3\times10^{10}$ & $4.69\times10^9$ & $3.1\times10^4$ & $7.1\times 10^{56}$ & $1.1\times 10^{57}$ & $1.2\times 10^{57}$ & $0.59$            & $0.92$ \\
\hline
M3e8-lowfg  & $3\times10^8$    & $1.2\times10^5$ & $34$           & $9.0\times 10^{50}$ & $1.2\times 10^{54}$ & $2.8\times 10^{56}$ & $3.2\times10^{-6}$ & $4.3\times10^{-3}$ \\
M1e9-lowfg  & $10^9$           & $7.1\times10^5$ & $200$          & $1.2\times 10^{52}$ & $7.1\times 10^{54}$ & $3.3\times 10^{56}$ & $3.6\times10^{-5}$ & $0.022$ \\
M3e9-lowfg  & $3\times10^9$    & $3.5\times10^6$ & $980$          & $1.2\times 10^{53}$ & $3.5\times 10^{55}$ & $4.4\times 10^{56}$ & $2.7\times10^{-4}$ & $0.080$ \\
M1e10-lowfg & $10^{10}$        & $2.2\times10^7$ & $6.2\times10^3$ & $1.7\times 10^{54}$ & $2.2\times 10^{56}$ & $7.2\times 10^{56}$ & $2.4\times10^{-3}$ & $0.31$ \\
M3e10-lowfg & $3\times10^{10}$ & $1.1\times10^8$ & $3.1\times10^4$ & $1.8\times 10^{55}$ & $1.1\times 10^{57}$ & $1.2\times 10^{57}$ & $0.015$           & $0.92$ \\
\hline
\hline
\end{tabular}
\caption{Parameters of TDE-powered AGN outflow simulations. First
  column shows model ID, the next four columns give the total mass,
  baryon fraction, gas mass and initial mass of central BH. The final
  three columns give the gas binding energy, the expected energy from
  SN explosions over the Hubble time, and the expected energy release
  by the AGN over the Hubble time.}
\label{table:params}
\end{table*}

\begin{figure*}
  \centering
    \includegraphics[trim = 0 0 0 0, clip, width=\textwidth]{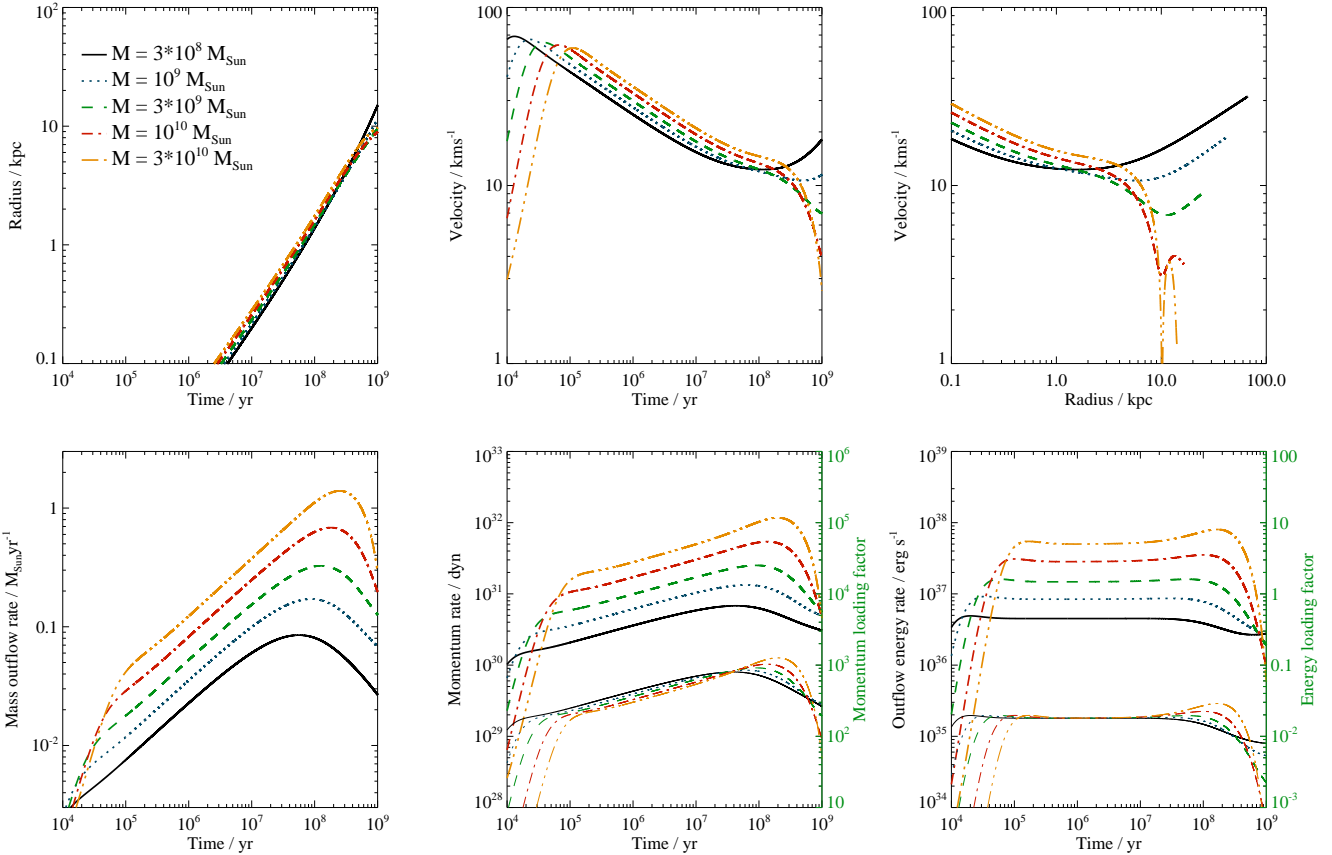}
  \caption{Outflow parameters in simulations with TDE-powered AGN,
    starting at $z_{\rm init} = 0.247$, with BH masses linearly
    proportional to stellar masses and baryon fraction $f_{\rm b} =
    0.16$. {\em Top left:} outflow radius against time. {\em Top
      middle:} outflow velocity against time. {\em Top right:} outflow
    velocity against radius. {\em Bottom left:} mass outflow rate
    against time. {\em Bottom middle}: outflow momentum rate (thick
    lines) and momentum loading factor $\dot{p}_{\rm out}c/\lbrace
    L_{\rm AGN}\rbrace$ (thin lines). {\em Bottom right:} outflow
    kinetic energy rate (thick lines) and energy conversion efficiency
    $\dot{E}_{\rm kin}/\lbrace L_{\rm AGN} \rbrace$ (thin lines). The
    five lines correspond to models with different total masses (see
    Table \ref{table:params} and text for details).}
  \label{fig:out_normal}
\end{figure*}

\begin{figure*}
  \centering
    \includegraphics[trim = 0 0 0 0, clip, width=\textwidth]{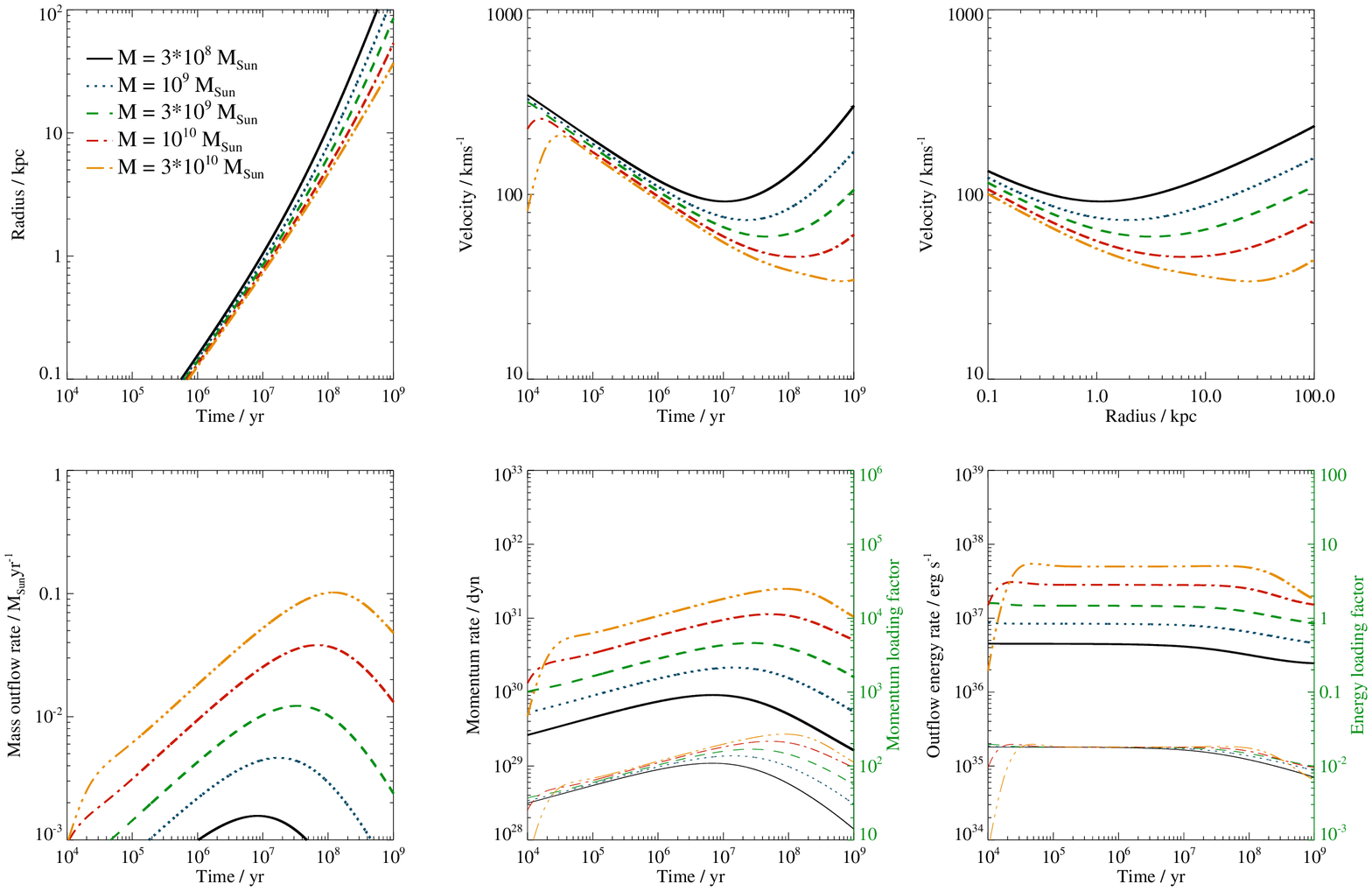}
  \caption{Outflow parameters in simulations with TDE-powered AGN,
    starting at $z_{\rm init} = 0.247$, with BH masses linearly
    proportional to stellar masses and gas mass equal to stellar
    mass. Panels and lines as in Figure \ref{fig:out_normal}.}
  \label{fig:out_lowgas}
\end{figure*}

I now investigate the propagation of outflows in spherical potentials
with parameters appropriate for dwarf galaxies. Given that a bursty
activity history caused by TDEs is equivalent, for the purposes of
outflow generation, to continuous activity with an appropriately
reduced AGN luminosity, I choose an average AGN luminosity dependent
only on its mass. Combining equations (\ref{eq:dmdt}) and
(\ref{eq:erad}), this luminosity is
\begin{equation} \label{eq:mdotaverage}
\langle L_{\rm AGN}\rangle = \eta \dot{M}_{\rm BH} c^2 = 4.2 \times
10^{39} \eta_{0.1}^{0.6} M_5^{0.353} m_*^{0.2} r_*^{0.6} {\rm erg \;
  s}^{-1}.
\end{equation}
I consider galaxies with five values of total mass: $M_{\rm tot} =
3\times 10^8, 10^9, 3\times 10^9, 10^{10}$ and $3\times 10^{10}
\msun$; the last mass exceeds that of the LMC \citep[$M_{\rm LMC} \leq
  3\times 10^{10} \msun$; ][]{vanderMarel2002AJ,
  Kallivayalil2006ApJ}. The present-day stellar mass $M_*$ is taken
from the abundance matching relation in \citet{Read2017MNRAS}, giving
$M_*\left(z=0\right) = 1.2\times 10^5, 7.1\times10^5, 3.5\times10^6,
2.2\times10^7$ and $1.1\times10^8 \msun$ for the five values of total
mass. I further assume that $M_*\left(z=0.247\right) =
M_*\left(z=0\right)$, i.e. that no significant star formation occurred
in these galaxies for the past 3 Gyr. This is approximately correct
for dwarf spheroidal and elliptical galaxies in the Local Group
\citep{Weisz2014ApJ}.

I consider two possible values of gas mass in the galaxy, leading to
10 models in total. In one group of models, I assume a baryon fraction
$f_{\rm b} = 0.16$ and take $M_{\rm gas} = f_{\rm b}M_{\rm tot} -
M_*$. This gives an upper limit to the gas mass, since gas removal by
stellar processes is not taken into account; however, some observed
galaxies have gas masses close to this limit \citep{Oh2015AJ,
  Kirby2017ApJ}. In the second case, I use $M_{\rm gas} = M_*$, which
represents a rather drastic case of gas removal, but agrees quite well
with the trend in observed dwarf galaxies. The resulting gas masses are given in
Table \ref{table:params}.

The initial black hole masses are selected using the black hole mass -
stellar mass relation \citep{Reines2015ApJ}: $M_{\rm BH} \simeq
10^{-3.55} M_*$. This results in extremely low black hole masses in
the first three cases, $M_{\rm BH,0} < 10^3 \msun$. However, it turns
out that even such tiny black holes are capable of driving outflows,
and if the actual black holes were larger, due to the expected rapid
growth at high redshift via BH mergers \citep{Devecchi2009ApJ} or
direct collapse \citep{Volonteri2008MNRAS}, the outflows would only be
more powerful.

Before presenting the numerical results, I note that it is possible to
estimate the expected TDE-powered AGN outflow effect by considering
the energy release by the AGN, the binding energy of the gas and the
supernova energy injection. These energy values are given in columns
5-7 of Table \ref{table:params}. For supernova energy injection, I
assume one supernova per $M_1 \sim 100 \msun$ of stars formed and $E_1
= 10^{51}$~erg per supernova. The last two columns of the table give
ratios $E_{\rm b}/E_{\rm AGN}$ and $E_{\rm SN}/E_{\rm AGN}$,
respectively. Given that the fraction of supernova energy communicated
to the ISM is $\lesssim 10\%$ \citep{Thornton1998ApJ, Walch2015MNRAS,
  Fierlinger2016MNRAS}, and the fraction of AGN energy communicated to
the ISM is $\sim 5\%$, it is clear that TDE-powered AGN outflows
should be more important than supernovae in the four smallest
galaxies, but probably not in the largest one. The total energy input
by stellar winds is similar to that of supernovae \citep{Chu2005ASPC,
  Voss2009A&A}, therefore the relative importance of AGN outflows to
stellar feedback is similar when AGN wind feedback is taken into
account. Similarly, AGN outflow energy is higher than the gas binding
energy in all five gas-poor and three smallest gas-rich models,
suggesting that TDE-powered outflows are probably more important in
these smallest galaxies. This can be understood by considering that
the average AGN luminosity scales as $M_{\rm BH}^{0.353}$, while BH
mass scales approximately as $M_{\rm tot}^{1.5}$, so $L_{\rm AGN}
\propto M_{\rm tot}^{0.53}$; on the other hand, the supernova energy
release scales as $M_* \propto M_{\rm tot}^{1.5}$, while the binding
energy of the gas scales as $M_{\rm tot}^2$. Therefore, the importance
of TDE-powered AGN outflows decreases with increasing galaxy mass.

The main results are presented in Figures \ref{fig:out_normal}
($f_{\rm b} = 0.16$) and \ref{fig:out_lowgas} ($M_{\rm gas} =
M_*$). In each figure, the six panels show, from left to right and top
to bottom, outflow radius against time, velocity against time,
velocity against radius, mass outflow rate against time, outflow
momentum rate and momentum loading factor $\dot{p}c/L_{\rm AGN}$
against time, and energy rate and energy conversion efficiency
$\dot{E}/L_{\rm AGN}$ against time. Note that the velocity and mass
outflow rate scales are different in the two figures. The AGN
luminosity used in these comparisons is the average long-term
luminosity $\lbrace L_{\rm AGN}\rbrace$; the actual luminosity during
tidal disruption events can be several orders of magnitude higher.

Outflows exist in all models considered here. This is not surprising:
assuming a perfectly adiabatic spherically symmetric system, the only
possibility of using the injected energy is by gas expansion. It is
more important to consider the properties of the outflows and check
whether they might be observable. Outflows in gas-rich simulations
expand with velocities $10 {\rm km/s} < v_{\rm out} < 100 {\rm km/s}$,
which is comparable to the velocity dispersions in real galaxies
(Figure \ref{fig:dgtdyn}, middle panel). Therefore such outflows would
be difficult to identify, since random gas motions would very easily
disrupt the outflow bubble. A large fraction of energy might escape
via low-density channels opened by gas turbulence, leading to collapse
of any outflow bubble that forms. However, the prolonged AGN energy
injection may result in a more radially-dominated velocity anisotropy
of gas motions in the galaxy. Formally, outflow radius becomes larger
than $r_{\rm eff} \sim 1$~kpc in $<10^8$~yr, i.e. the effects of AGN
outflows can manifest rather quickly, in a few dynamical times of the
galaxy. The mass outflow rate can reach $0.1-1 \msun$~yr$^{-1}$, a
much higher value than typical star formation rates
\citep{McGaugh2017ApJ}. Therefore gas may be gradually pushed out from
the central parts of the galaxy; this can potentially quench star
formation to some extent.

In the simulations with low gas density, outflows expand much more
rapidly, with $v_{\rm out} > 100$~km/s in the central parts of the
galaxy, and all outflows accelerate after passing through the highest
circular velocity region of the potential. This should result in
outflows that are detectable and might be significant in
redistributing the gas in the galaxy. Importantly, in all simulations,
outflows reach $r_{\rm out} = 100$~kpc in several Gyr or less,
i.e. gas can be removed outside the virial radius of the galaxy. The
mass outflow rate is, naturally, much lower than in the high gas
density simulations, but even the $10^{-3} \msun {\rm yr}^{-1} <
\dot{M}_{\rm out} < 0.1 \msun {\rm yr}^{-1}$ rates are comparable to
or larger than typical star formation rates in similar galaxies
\citep{McGaugh2017ApJ}. The integrated mass outflow over the course of
the simulation is comparable to the initial gas mass in the
galaxy. Therefore, even though the outflow itself might be weak and
difficult to detect, perhaps better described as an expanding galactic
atmosphere, it might still have a noticeable effect on the host
galaxy.

The momentum and kinetic energy rates of outflows in both cases are
modest. The momentum loading factors can be very large due to the low
outflow velocities ($\dot{p}c/\lbrace L_{\rm AGN}\rbrace > 10$), but
would be much lower if the outflow was detected during a period of
nuclear activity following a TDE, because then the momentum rate would
be compared with instantaneous, rather than average, AGN
luminosity. The same is true of the energy conversion
efficiency. Therefore even if an outflow signature is detected in a
dwarf galaxy, it might be difficult to connect it to sporadic nuclear
activity rather than stellar processes. This has previously been noted
in the context of fossil outflows in large galaxies
\citep{King2011MNRAS, Fluetsch2018arXiv, Nardini2018MNRAS}.

\section{Discussion} \label{sec:discuss}

\subsection{Current and seed black hole masses}

The black hole masses used in the calculations of outflow properties
are linearly proportional to galaxy stellar masses. If, instead, a
lower limit $M_{\rm BH,min} = 10^3-10^4 \msun$ were imposed on the
seed SMBH mass, one would expect somewhat stronger outflows in
galaxies with $M_* \simlt 5\times10^6 \msun$. The outflow velocity
scales approximately as $v_{\rm out} \propto L_{\rm AGN}^{1/3} \propto
M_{\rm BH}^{0.12}$, while the mass flow rate $\dot{M}_{\rm out}
\propto L_{\rm AGN}^{1/2} \propto M_{\rm BH}^{0.18}$; the latter
scaling is stronger than predicted by \citet{Zubovas2012ApJ} because
of the radial increase of effective $\sigma$ in an NFW potential. The
outflow kinetic power then scales as $\dot{M}_{\rm out} v_{\rm out}^2
\propto M_{\rm BH}^{0.42}$, i.e. the kinetic power with massive black
hole seeds could be as much as $7-20$ times greater than in galaxies
with stellar seeds.

Such outflows would be more easily detectable in the local Universe
and might allow checking the hypotheses of SMBH formation. On the
other hand, even stellar-mass black holes should grow to $M_{\rm f} >
10^3 \msun$ by $z=0$ (see eq. \ref{eq:m0mf}), so observations of more
distant dwarf galaxies are required in order to be able to distinguish
between models of seed black hole masses. As a rough estimate, if we
assume that the black hole mass can be determined to within an order
of magnitude based on outflow properties, the existence of seed black
hole masses $M_{\rm seed} \simgt 10^4 \msun$ can be checked by
observing dwarf galaxies at $t \simlt 10$~Gyr after the Big Bang ($z
\simgt 0.33$), since after that time, a stellar black hole with
initial mass $M_0 = 10 \msun$ will have grown to $M_{\rm f} > 10^3
\msun$. Checking the existence of black holes with $M_{\rm seed}
\simgt 10^3 \msun$ would require observing dwarf galaxies at $t \simlt
2.1$~Gyr after the Big Bang ($z \simgt 3.1$). Although large numbers
of dwarf galaxies with $M_* > 10^7 \msun$ are known beyond the Local
Volume \citep[e.g.][]{Mezcua2016ApJ, Mezcua2018MNRAS}, the faintest
dwarfs are so far only found in the Local Universe
\citep{Lee2017ApJ}. Therefore constraining black hole seed masses
using outflow properties is unlikely now, although might become
possible in the near future, especially with next generation
instruments.

The growth of black holes via TDEs in the smallest galaxies would lead
to those black holes becoming overmassive compared to the predictions
of black hole - stellar mass correlations. There is a hint that this
relation flattens at low masses \citep[fig. 9]{Reines2015ApJ}, but
dynamical measurements of black hole masses in dwarf galaxies are
needed before this flattening can be confirmed. If it is, it would
provide strong support to the hypothesis that TDEs are an important
source of black hole growth in dwarf galaxies.

Another interesting issue is that in the smallest galaxies considered
here, the central black hole mass is of the same order as stellar
black holes can be, i.e. $M_{\rm BH} < 100 \msun$. Since even such a
small object may create significant outflows in the host galaxy, it is
worth considering that several stellar black holes, each powered by
independent TDEs, may have an even larger effect on the smallest dwarf
galaxies, efficiently driving gas out of the galaxy and leading to the
very high observed mass-to-light ratios \citep{Gilmore2007ApJ,
  Wolf2010MNRAS}. In this sense, stellar black holes may have a
similar effect to ultra-luminous X-ray sources \citep[ULXs;
  cf.][]{King2001ApJ}, except that their feedback would be distributed
in time throughout the age of the galaxy, rather than concentrated
around episodes of star formation.

In dwarf galaxies, especially the smallest ones, massive black holes
are not necessarily located in the centre: the gravitational potential
is so shallow, and the mass ratio between the central object and
individual stars so low, that significant black hole wandering can
occur \citep{Bellovary2018arXiv}. This process would lead to highly
asymmetric disturbances to galactic gas after each TDE and associated
activity episode. This may perhaps explain some puzzling features in
dwarf galaxies, e.g. the wide holes in the gas distributions in Leo A
and Aquarius \citep{Hunter2012AJ}.

\subsection{Observability of TDE-driven outflows}

The energies released by TDEs in dwarf galaxies are typically higher than
gas binding energies or supernova energy release (see Table
\ref{table:params}), even accounting for the $5\%$ coupling efficiency
of this energy to the ISM. However, outflows produced by TDE-powered
AGN don't always break out of galaxies and will not generally be
easily detectable due to their low velocities (see Section
\ref{sec:outflow_results} and Figures \ref{fig:out_normal} and
\ref{fig:out_lowgas}). Therefore, the AGN energy input may manifest
differently, for example, by creating a slowly expanding gaseous
`atmosphere' or by increasing gas turbulence in the galaxy. These
effects can be detected by observing gas kinematics and density
profiles. In the long run, this process may significantly decrease
the star formation efficiency in dwarf galaxies. This may explain the
low observed star formation efficiency \citep{McGaugh2017ApJ}. In
galaxy samples selected by dynamical mass, differences in stellar mass
would then correlate with presence of central massive black holes.

On shorter timescales, disturbances of dwarf galaxy ISM should persist
for at least an order magnitude longer than the AGN episode that drove
the outflow \citep{King2011MNRAS}. Therefore, we can expect $\sim10\%$
of dwarf galaxies that harbour massive black holes to show
disturbances in their ISM. This number can be used to estimate SMBH
occupancy fraction in dwarf galaxies. More detailed numerical
simulations, which are beyond the scope of this paper, would help
predict the expected effects in more detail and could be tested with
spatially resolved dwarf galaxy observations, which will be provided
by, e.g., the Euclid space observatory \citep{Laureijs2011arXiv}.

As each TDE lasts for only $t \lesssim 100$~yr, the effects of
individual activity episodes may be visible to some extent in the
galaxy. The outer edge of the expanding atmosphere or zone of
increased turbulence (see above) moves with the low average velocity
(Section \ref{sec:outflow_results}), but each individual episode can
create a faster shell propagating inside this zone. The speed on
individual shells is $v_{\rm ind}/v_{\rm out} \simeq \left(L_{\rm
  AGN}/\langle L_{\rm AGN} \rangle\right)^{1/3} \simeq f_{\rm
  AGN}^{-1/3} \sim 5$. In almost all the cases considered in this
paper, this leads to outflows moving with velocities $v_{\rm ind} >
100$~km~s$^{-1}$, and sometimes with $v_{\rm ind} >
10^3$~km~s$^{-1}$. Such individual outflow shells would be easily
detectable. The most important caveat here is that the properties of
the shell motion do not depend strongly on the AGN feeding source,
therefore AGN fed by gas streams should cause essentially identical
outflows. The presence of gas moving with high radial velocities is
therefore evidence of an AGN episode in the recent past, rather than
evidence of the AGN feeding mechanism. The presence of multiple shell
moving with different radial velocities, however, may signal that
there have been several individual AGN episodes in the recent past,
which is more likely if the episodes are triggered by TDEs than if
they are fed by large gas streams. Detection of individual shells
requires spatially-resolved spectra of dwarf galaxies, but this may be
possible in the near future.

Shells expanding through the turbulent ISM of the dwarf galaxy would
be subject to various instabilities, most notably the Rayleigh-Taylor
and Richtmyer-Meshkov instabilities due to impacts with the
surrounding ISM, and Vishniac instability \citep{Vishniac1983ApJ} due
to density increase and self-gravity. The compaction of ISM and the
increased mixing of material with different orbital energies and
angular momenta can in fact lead to enhancement of SMBH accretion via
gas streams \citep{Dehnen2013ApJ}. In this way, TDE-powered AGN
outflows can kick-start the process of SMBH growth in dwarf
galaxies. Magnetic fields can prevent the growth of instabilities
\citep{Diehl2008ApJ}, but are unlikely to be very efficient, since
dwarf galaxies tend to have weaker magnetic fields than larger ones
\citep{Chyzy2011A&A}.

Finally, the repeated shell expansion and contraction due to
individual TDE-powered AGN episodes may lead to a relaxation of the
central dark matter cusp, as seen in simulations of repeated supernova
feedback \citep{Pontzen2012MNRAS,Governato2012MNRAS}. This process,
therefore, may help solve the cusp-core problem in dwarf galaxies, as
suggested by \citet{Silk2017ApJ}.

\subsection{Enhancement of star formation rates}

In the gas-rich galaxy simulations, outflow pressure peaks at $\sim
10^4 - 10^5$~yr, at a few times $10^7$~K~cm$^{-3}$, and then decreases
as a power law with time. Assuming an ISM pressure $P_{\rm ISM}/k_{\rm
  B} = 10^5$~K~cm$^{-3}$ \citep{Young2001ASPC}, the outflow pressure
exceeds the ISM pressure for a few Myr. Therefore, at very early
times, outflowing gas might compress denser clumps in the ISM and
enhance the star formation rate \citep{Silk2005MNRAS,
  Zubovas2014MNRASc, Zubovas2016MNRASb}. This should not be a
significant effect: even assuming that the whole star-forming ISM is
affected and that the star formation rate increases linearly with
pressure enhancement, the expected SFR increase will not exceed a
factor $R_{\rm SFR} \sim 100$. Given that the specific star formation
rate in galaxies is $<10^{-8}$~yr$^{-1}$, even at high redshift
\citep{Behroozi2013ApJ}, the triggered star formation can only lead to
a stellar mass increase of $10^{-8}$~yr$^{-1} \cdot R_{\rm SFR} \cdot
10^6$~yr$\sim 1$ times the natural star formation. In other words, the
outflow-triggered star formation can be responsible for at most $50\%$
of the star formation in the first 100~Myr of the dwarf galaxy's
evolution. Detecting such differences would require very precise
determination of the star formation history and initial conditions of
the dwarf galaxy.

\subsection{Other modes of feedback} \label{sec:other_modes}

One more complication to understanding the growth on SMBHs in dwarf
galaxies is feeding by gas streams. Even though it is unlikely to
contribute much to black hole growth (see Section \ref{sec:feed}),
this process can still occur occasionally, feed the black hole and
affect the host galaxy significantly. \citet{Dashyan2018MNRAS}
investigated the possibility of AGN outflows in dwarf galaxies while
treating the AGN luminosity as a free parameter, and found that AGN
feedback can be more powerful than supernova feedback, if the AGN are
fed at high Eddington ratios and/or if the supernova wind coupling
efficiency is low. These results are qualitatively similar to those
presented here, since TDE-powered AGN can also have phases of high
Eddington ratio accretion. One drawback of that study is that the
$p$d$V$ work done by the expanding gas, which accounts for $\sim 1/3$
of the injected AGN wind energy \citep{Zubovas2012ApJ}, was not
included in the outflow energy equation (their eq. 7), so the actual
AGN outflow powers should be smaller than presented there.

It is now established that TDEs can launch jets from the accretion
disc \citep{Levan2011Sci, Burrows2011Natur, Cenko2012ApJ,
  Komossa2015JHEAp}. The luminosity of the jet can, at least
initially, be much higher than that of the disc
\citep{Piran2015MNRAS}. However, the duration of peak jet activity is
of order of weeks or months \citep{Mimica2015MNRAS}, much less than
the lifetime of the accretion disc built up during a TDE (see Section
\ref{sec:tde_rates}). Therefore the importance of the jet on long
timescales is most likely small. In addition, the major effects of
jets should be confined to sub-parsec scales and manifest on several
year timescales \citep{Giannios2011MNRAS}. Even if the jet is
energetically important, its overall effect may be similar to that of
a wide-angle wind-driven outflow, since the jet may inflate a bubble
in the host galaxy and affect its ISM throughout the host
\citep[e.g.,][]{Gaibler2012MNRAS}.

\subsection{Validity of TDE rates}

The TDE rates used in this work are theoretical estimates which are
based on rather idealized assumptions regarding the distribution of
stars in the galaxy. Some processes, for example the presence of
circumnuclear stellar rings and eccentricity oscillations therein, may
raise the TDE rate, possibly significantly \citep{Madigan2018ApJ}.
Other processes might reduce the rate: for example, the TDE rate
extrapolated from the results of \citet{Stone2016MNRAS} to the
smallest galaxies results in most of the stellar population being
disrupted over the Hubble time. This is unlikely to be the case, and
in such galaxies, the calculation of TDE rates should take into
account the finite amount of available stars, leading to a lower
estimate \citep[for example, as in numerical simulations
  by][]{Brockamp2011MNRAS}. Nevertheless, it is interesting to note
that the decrease of stellar mass due to TDEs might contribute to the
very high mass-to-light ratios in the smallest galaxies
\citep{Gilmore2007ApJ, Wolf2010MNRAS, Read2017MNRAS}.

Observational estimates of the true rates of tidal disruption events
are also difficult to make. Many theoretical predictions might be
overestimates by as much as an order of magnitude \citep[see
  discussion in][]{Stone2016MNRAS}. However, a broad range
$\dot{N}_{\rm TDE} \sim 10^{-5} - 10^{-4}$~yr$^{-1}$ per galaxy, given
in a review by \citet{Komossa2015JHEAp}, falls within the range used
in this work for $M_{\rm BH} > 1.9 \times 10^4 \msun$, which is
consistent with the mass range for which observational constraints are
available. The true rate might even be somewhat higher than these
observations suggest, for two reasons. Highly super-Eddington TDEs may
not follow the canonical $t^{-5/3}$ luminosity profile due to
accretion being limited by viscous transport \citep{Lin2017NatAs,
  Wu2018MNRAS} and therefore might not always be identified as
such. Secondly, TDEs around the smallest black holes may be more
difficult to detect because their peak luminosity is smaller
\citep{DeColle2012ApJ}, and because the probability of prompt
emission, with a different light curve, resulting from direct impact
by the star becomes non-negligible, since the tidal disruption radius
becomes only a few times the stellar radius.

Overall, if the actual TDE rate is significantly lower than that
calculated by \citet{Stone2016MNRAS}, then TDE-powered AGN and their
associated outflows are much rarer than predicted here, and would be
unable to explain the observed population of AGN in dwarf galaxies. If
the actual TDE rate is higher, then the predicted AGN fraction in
dwarf galaxies is also higher and implies that not all dwarf galaxies
harbour massive central black holes. In either case, future
theoretical work and better observational constraints will help
determine the frequency of TDEs and the occupation fraction of massive
black holes in dwarf galaxies.

\section{Summary and conclusions} \label{sec:concl}

I have investigated the possibility of powering AGN by tidal
disruption events and of such AGN driving outflows in dwarf
galaxies. The results, based on analytical arguments and numerical
calculations of outflow properties, are the following:

\begin{itemize}
\item TDEs can feed supermassive black holes in dwarf galaxies,
  producing mass inflow rates large enough for AGN episodes to last,
  in total, for a significant fraction ($\simgt1\%$) of Hubble time.

\item Accounting for the viscous timescale of matter infall suggests
  that each dwarf galaxy hosting a SMBH can be active for an even
  larger fraction of the Hubble time.

\item The total energy released by TDE-powered accretion episodes in a
  galaxy with $10^5 \msun < M_* < 10^8 \msun$ is larger than the total
  energy released by supernovae or the binding energy of the gas in
  the same galaxy.

\item Outflows driven by TDE-powered AGN can break out of the smallest
  gas-rich dwarf galaxies and out of most gas-poor dwarf galaxies.

\item The velocities of outflows driven by TDE-powered AGN in gas-rich
  dwarf galaxies are small, making these outflows difficult to detect;
  instead, their effects might be observable as increased turbulence
  and extended gas distribution.
  
\end{itemize}

As more data of AGN in dwarf galaxies, as well as spatially resolved
dwarf galaxies, become available, the predictions presented here may
be tested and used to constrain the occupancy fraction of massive
black holes in dwarf galaxies, as well as the actual effects they have
on the evolution of their hosts.

\section*{Acknowledgments}

I thank Chris Nixon for comments and suggestions during the
preparation of this manuscript, and the anonymous referee for
insightful comments that helped put the findings presented here in
their proper context. This research was funded by a grant
(No. LAT-09/2016) from the Research Council of Lithuania.


\end{document}